\documentclass[prl,nofootinbib,elsart]{revtex4}

\usepackage{graphicx}
\usepackage{latexsym}
\usepackage{amsmath}

\begin{document}
\title{Two-dimensional Noncommutative atom Gas with Anandan
interaction}

\author{Xiaomin Yu$^a$}

\author{Kang Li$^b$
\footnote{Corresponding email: kangli@hznu.edu.cn}}

\affiliation{$^{a)}$Dean's Office, Hangzhou Dianzi University,
Hangzhou, 310018, China}

\affiliation{$^{b)}$Department of Physics, Hangzhou Normal
University, Hangzhou, 310036, China}

\begin{abstract}
Landau like quantization of the Anandan system in a special
electromagnetic field is studied. Unlike the cases of the AC
system and the HMW system, the torques of the system on the
magnetic dipole and the electric dipole don't vanish. By
constructing Heisenberg algebra, the Landau analog levels and
eigenstates on commutative space, NC space and NC phase space are
obtained respectively. By using the coherent state method, some
statistical properties of such free atom gas are studied and the
expressions of some thermodynamic quantities related to revolution
direction are obtained. Two particular cases of temperature are
discussed and the more simple expressions of the free energy on
the three spaces are obtained. We give the relation between the
value of $\sigma$ and revolution direction clearly, and find
Landau like levels of the Anandan system are invariant and the
levels between the AC system and the HMW system are interchanged each other
under Maxwell dual transformations on the three spaces. The two
sets of eigenstates labelled by $\sigma$ can be related by a
supersymmetry transformation on commutative space, but the
phenomenon don't
occur on NC situation. We emphasize that some results relevant to Anandan interaction are suitable for the cases of  AC interaction and HMW interaction under special conditions.\\



\end{abstract}

\maketitle

\section{1. Introduction}
There are many papers to study the topological and geometrical
effects of charged and neutral particles in the presence of
electromagnetic fields in their quantum dynamics. One well known
topological effect is the Aharonov-Bohm (AB) effect which shows
that a quantum charge circulating a magnetic flux line accumulates
a quantum topological phase and demonstrates the physical
significance of magnetic vector potential.\cite{AB} The effect can
be observed by using matter-wave interferometry. Another well
known quantum phase is the Aharonov-Casher (AC) phase acquired by
a neutral particle with a non-zero magnetic dipole moment,
circulating a straight line of charge \cite{AC}, which is a
non-dispersive quantum geometrical phase \cite{Badurek} and was
observed experimentally in a neutron interferometer \cite{Cimmino}
and in a neutral atomic Ramsey interferometer \cite{Sangster}. Its
Maxwell dual phase is the He-Mckellar-Wilkens (HMW) phase which
implies that a neutral particle with a non-zero electric dipole
moment moving around a line of magnetic monopoles would accumulate
a quantum phase.\cite{HMW} Wei, Han and Wei proposed a practical
experimental configuration to test this effect \cite{WHW}, and
Dowling et al proposed two other experimental schemes for it
\cite{Dowling}. Recently, another non-dispersive quantum
geometrical phase was proposed by Anandan. It describes that a
neutral particle with permanent electric dipole moment and
non-vanishing magnetic dipole moment in the presence of external
magnetic and electric fields in the relativistic and
non-relativistic case accumulates a
quantum phase.\cite{AnandanR}-\cite{Furtado} This phase is called the Anandan phase.\\

The interaction of the magnetic field with a charged particle in
two dimensions plays an important role in landau quantization and
quantum \emph{Hall effect}. Inspired by the work of Paredes
\emph{et al.}, proving the existence of anionic excitation in
rotating Bose-Einstein condensates \cite{Paredes}, Ericsson and
Sj$\ddot{o}$qvist used the AC interaction and developed an analog
of Landau quantization which provide the possibility of an atomic
quantum \emph{Hall effect} \cite{Ericsson}. Following these steps,
Ribeiro \emph{et al.} made use of the HMW interaction to generate
another analog of
Landau quantization.\cite{Ribeiro}\\

Recently, the study of physics effects on noncommutative (NC)
space and noncommutative phase space has attracted much
interest.\cite{SW}-\cite{nair} There are many researches in NC
quantum mechanics such as  the AB effect
\cite{Chaichian1}-\cite{klisd}, the AC effect \cite{Mirza,Mirza1},
the HMW effect \cite{Ribeiro,Dulat}, Landau levels
\cite{Horvathy1}-\cite{Gamboa}, quantum \emph{Hall effect}
\cite{Susskind}-\cite{Basu}, and so on. Unlike the case in usual
quantum mechanics, Passos \emph{et al.} have demonstrated that the
AC phase, the HMW phase and the Anandan phase are quantum
geometric dispersive phases on NC space and NC phase
space.\cite{Passos} They have also obtained noncommutative Landau
like quantization with the AC interaction and the HMW interaction
respectively.\cite{RibeiroE} In this letter, we will use the
Anandan interaction for another analog of Landau quantization in
the non-relativistic limit on commutative space (C), NC space and
NC phase space. Unlike the Landau like quantization of the AC
system and the HMW system, both torques of the Anandan system on
the magnetic dipole and the electric dipole don't vanish. We use
the coherent state method to give out the expressions of some
thermodynamic quantities related to revolution direction for the
free atom gas with the Anandan interaction and discuss the free
energy of the system on two particular cases of temperature on the
three spaces. By the description of atom orbits, we give the
relation between the value of $\sigma$ and revolution direction.
We find that Landau like levels of the Anandan system are
invariant and the levels between the AC system and the HMW system
are interchanged each other under Maxwell dual transformations on the three
spaces. We think the levels of the AC system and the HMW system
are similar under the same revolution direction which is different
from the point of Ribeiro \emph{et al.}\cite{Ribeiro} and analyze
their mistake from  the angles of revolution direction and Maxwell
duality. Some difference between commutative space and NC
situation is also discussed from the supersysmmetry
transformation. Some results from the Anandan system can be
restricted to the ones of the AC system and the HMW system.\\

This paper is organized as follows: In section 2 and 3 the Landau
like quantization of a neutral atom with permanent electric and
magnetic dipole moments in a special electromagnetic field on
commutative space and NC situation is studied. In section 4 some
thermodynamic properties of such free atom gas are studied and two
particular cases of temperature are discussed on commutative
space, NC space and NC phase space. Some results are restricted to
the AC system and the HMW system. In section 5 we study the
relation between the value of $\sigma$ and revolution direction by
the coherent state, discuss the levels of the Anandan system, the
AC system and the HMW system by Maxwell duality, and analyze the
mistake of Ribeiro \emph{et al.} by the angles of revolution
direction and Maxwell duality. In section 6, by super-symmetric study,  we
give out the difference of the system between commutative space
and NC situation. Finally, in section 7 we present our
conclusions.\\
\section{2. Landau levels analog for neutral atoms on commutative space}
In the non-relativistic limit, the Hamiltonian for a neutral
spin-half atom possessing a non-zero electric dipole moment
$\textbf{d}$ and a non-zero magnetic dipole moment $\textbf{u}$,
moving in an external electromagnetic fields, can be described by
the Anandan Hamiltonian \cite{Passos},
\begin{equation}
\label{Eq:Rep.4}
\begin{array}{ll}
H={1\over 2m}[\textbf{P}-c^{-2}(\textbf{u}\times
\textbf{E})+c^{-2}(\textbf{d}\times
\textbf{B})]^2-\frac{u\hbar}{2mc^2}\nabla\cdot
\textbf{E}+\frac{d\hbar}{2mc^2}\nabla\cdot \textbf{B}
\end{array}\end{equation}
where the terms of $O(E^2)$ and $O(B^2)$ are neglected.
Considering the quantum dynamics of a particle, the Hamiltonian in
fact contains two other physical situations, the AC effect
($u\neq0$ and $d=0$) and the HMW effect ($u=0$ and $d\neq0$). In
the present problem, the kinematic momentum is
$\Pi=-i\hbar\nabla-c^{-2}(\textbf{u}\times
\textbf{E})+c^{-2}(\textbf{d}\times \textbf{B})$ and the Anandan
vector potential is defined as $\textbf{A}=(\textbf{u}\times
\textbf{E})-(\textbf{d}\times \textbf{B})$. Thus the associated
field strength is $\textbf{B}_{eff}=\nabla\times \textbf{A}$. Here
we consider the atom moves in the $x-y$ plane, and its magnetic
dipole moment as well as electric dipole moment are in the
$z-$direction. At the same time, the electric field configuration
$E=\frac{\rho_e}{2}r\hat{e}_\phi$ and the magnetic field
configuration $B=\frac{\rho_m}{2}r\hat{e}_\phi$ mentioned in
Ref.\cite{Ribeiro} are still used. Obviously, $\textbf{B}_{eff}$
is uniform and the conditions for electrostatics and
magnetostatics are satisfied. But unlike the cases of Landau
analogous quantization with the AC interaction and the HMW
interaction, both torques on the magnetic dipole and the electric
dipole don't vanish, because the atom with momentum
$\langle\Pi\rangle$ sees an effective magnetic field
$\textbf{B}'\cong \textbf{B}+\textbf{v}\times \textbf{E}/c$ and an
effective electric field $\textbf{E}'\cong
\textbf{E}-\textbf{v}\times \textbf{B}/c$ in its own reference
frame. The Hamiltonian (1) in such dipole-field configuration can
be written as
\begin{equation}
\label{Eq:Rep.4}
\begin{array}{ll}
H=\frac{1}{2m}[(p_{x}+\frac{u\rho_{e}-d\rho_{m}}{2c^2}y)\hat{\textbf{e}}_{x}+(p_{y}-\frac{u\rho_{e}-d\rho_{m}}{2c^2}x)\hat{\textbf{e}}_{y}]^2-\frac{(u\rho_{e}-d\rho_{m})\hbar}{2mc^2}
\end{array}\end{equation}
Further, the Hamiltonian can be expressed as
\begin{equation}
\label{Eq:Rep.4}
\begin{array}{ll}
H=\frac{1}{2m}({p_x}^2+{p_y}^2)+\frac{1}{8}m\omega^2(x^2+y^2)+\frac{\omega}{2}(p_xy-p_yx)-\frac{\omega\hbar}{2}
\end{array}\end{equation}
with the cyclotron frequency $\omega=\sigma\frac{\mid
u\rho_{e}-d\rho_{m}\mid}{mc^2}=\frac{ u\rho_{e}-d\rho_{m}}{mc^2}$
where $\sigma=\pm1$. Thus, the cyclotron frequencies of the AC
system and the HMW system are $\omega_{AC}=\sigma\frac{\mid
u\rho_{e}\mid}{mc^2}=\frac{ u\rho_{e}}{mc^2}$ and
$\omega_{HMW}=\sigma\frac{\mid d\rho_{m}\mid}{mc^2}=\frac{
-d\rho_{m}}{mc^2}$, respectively. If we introduce the operators,
\begin{equation}
\label{Eq:Rep.4}
 \begin{array}{ll}
a_x=\sqrt{{m|\omega|}\over 8\hbar}(x+{{2\emph{i}p_x}\over
m|\omega|})-i\sqrt{{m|\omega|}\over 8\hbar}(y+{{2\emph{i}p_y}\over
m|\omega|}),\hspace{2cm} a^+_x=\sqrt{{m|\omega|}\over 8\hbar}(x-{{2\emph{i}p_x}\over m|\omega|})+i\sqrt{{m|\omega|}\over 8\hbar}(y-{{2\emph{i}p_y}\over m|\omega|}),                         \\~~\\
a_y=i\sqrt{{m|\omega|}\over 8\hbar}(x+{{2\emph{i}p_x}\over
m|\omega|})-\sqrt{{m|\omega|}\over 8\hbar}(y+{{2\emph{i}p_y}\over
m|\omega|}),\hspace{2cm} a^+_y=-i\sqrt{{m|\omega|}\over
8\hbar}(x-{{2\emph{i}p_x}\over m|\omega|})-\sqrt{{m|\omega|}\over
8\hbar}(y-{{2\emph{i}p_y}\over m|\omega|}),\\
\end{array}
\end{equation}
where they satisfy Heisenberg algebraic relations
\begin{equation}
\label{Eq:Rep.4}
 \begin{array}{ll}
~&~[{a}_i, {a}_j]=0,~[{a}_i^+, {a}_j^+]=0,[{a}_i,
{a}_j^+]=\delta_{ij},\hspace{0.5cm}i,j=x,y,
\end{array}
\end{equation}
the Hamiltonian becomes
\begin{equation}
\label{Eq:Rep.4}
 \begin{array}{ll}
H=\frac{\hbar|\omega|}{2}({a}^+_x{a}_x+{a}^+_y{a}_y+1)+\frac{\hbar\sigma|\omega|}{2}({a}^+_y{a}_y-{a}^+_x{a}_x)-\frac{\sigma|\omega|\hbar}{2}.
\end{array}
\end{equation}
Therefore, the eigenvalues of $H$ are
\begin{equation}
\label{Eq:Rep.4}
 \begin{array}{ll}
E_{n_x,n_y}=\frac{\hbar|\omega|}{2}(n_x+n_y+1)+\frac{\hbar\sigma|\omega|}{2}(n_y-n_x)-\frac{\sigma|\omega|\hbar}{2},
\end{array}
\end{equation}
where non-negative integers ${n}_x,{n}_y$ are the eigenvalues of
the number operators $a^+_xa_x$, $a^+_ya_y$, respectively. The
corresponding eigenstates are
\begin{equation}
|{n}_x,{n}_y \rangle={1\over
\sqrt{{{n}_x}!{{n}_y!}}}({a^+_x})^{{n}_x}({a^+_y})^{{n}_y}| 0,0
\rangle,\label{h1}
\end{equation}
where $| 0,0 \rangle$ is the vacuum of $H$. Thus, the levels for
the AC system and the HMW system are
\begin{equation}
E^{AC}_{n_x,n_y}=\frac{\hbar|\omega_{AC}|}{2}(n_x+n_y+1)+\frac{\hbar\sigma|\omega_{AC}|}{2}(n_y-n_x)-\frac{\sigma|\omega_{AC}|\hbar}{2}
\end{equation}
and
\begin{equation}
E^{HMW}_{n_x,n_y}=\frac{\hbar|\omega_{HMW}|}{2}(n_x+n_y+1)+\frac{\hbar\sigma|\omega_{HMW}|}{2}(n_y-n_x)-\frac{\sigma|\omega_{HMW}|\hbar}{2}
\end{equation}
respectively.
\section{3. Landau levels analog for neutral atoms on NC phase space and NC space}
In order to maintain the Bose-Einstein statistics in NC quantum
mechanics, non-commutativity of both space-space and
momentum-momentum may be necessary.\cite{nair} We need to consider
physical problems on this space, NC phase space, where the
coordinates $\hat{x}_i$ and momentums $\hat{p}_i$ satisfy the
following relations:
\begin{equation}
\label{Eq:nmr}
 \begin{array}{l}
~[\hat{x}_{i},\hat{x}_{j}]=i\Theta_{ij}, \hspace{1cm}
[\hat{p}_{i},\hat{p}_{j}]=i\bar{\Theta}_{ij},\hspace{1cm}
[\hat{x}_{i},\hat{p}_{j}]=i \hbar\delta_{ij},
\end{array}
\end{equation}
where the elements of the antisymmetric matrices $\{\Theta_{ij}\}$
and $\{\bar{\Theta}_{ij}\}$ are very small and represent
space-space non-commutativity and momentum-momentum
non-commutativity. The static Schr$\ddot{o}$dinger equation on NC
phase space is usually expressed as
\begin{equation}\label{h1}
 H(x,p)\ast\psi = E\psi,
\end{equation}
where $H(x,p)$ is the Hamiltonian of the usual quantum system and
a star product (Moyal-Weyl product) is defined by
\begin{equation}
(f  \ast g)(x,p) = e^{ \frac{i}{2\alpha^2}
 \Theta_{ij} \partial_i^x \partial_j^x+\frac{i}{2\alpha^2}\bar{\Theta}_{ij} \partial_i^p
 \partial_j^p}
 f(x,p)g(x,p)  = f(x,p)g(x,p)
 + \frac{i}{2\alpha^2}\Theta_{ij} \partial_i^x f \partial_j^x g\big|_{x_i=x_j}
 + \frac{i}{2\alpha^2}\bar{\Theta}_{ij} \partial_i^p f \partial_j^p g\big|_{p_i=p_j}.
\end{equation}
By replacing the star product by the general ordinary product and
shifting coordinates ${x_i}$ and momentums ${p_i}$ with
\cite{{CFZ}}
\begin{equation}
\label{Eq:Rep.4}
 \begin{array}{ll}
 {x}_{i}~\longrightarrow~\hat{x}_{i}&= \alpha x_{i}-\frac{1}{2\hbar\alpha}\Theta_{ij}p_{j},\\
 ~&~\\
 {x}_{i}~\longrightarrow~\hat{p}_{i}&=\alpha p_{i}+\frac{1}{2\hbar\alpha}\bar{\Theta}_{ij}x_{j},
\end{array}
\end{equation}
the Schr$\ddot{o}$dinger equation can be written as
\begin{equation}\label{GBShift}
H( \alpha x_{i}-\frac{1}{2\hbar\alpha}\Theta_{ij}p_{j}, ~ \alpha
p_{i}+\frac{1}{2\hbar\alpha}\bar{\Theta}_{ij}x_{j})\psi = E\psi.
\end{equation}
Here $\alpha$ is a scaling constant related to the
non-commutativity of phase space and satisfies the relation
$\theta\bar{\theta}=4\alpha^2(1-\alpha^2)$\cite{Likang}. So the
Hamiltonian of the system on NC phase space can be written as
\begin{equation}
\label{Eq:Rep.4}
\begin{array}{ll}
H=\frac{1}{2m}[((\alpha
p_{x}+\frac{1}{2\hbar\alpha}\bar{\theta}y)+\frac{u\rho_{e}-d\rho_{m}}{2c^2}(\alpha
y+\frac{1}{2\hbar\alpha}\theta
p_{x}))\hat{\textbf{e}}_{x}+((\alpha
p_{y}-\frac{1}{2\hbar\alpha}\bar{\theta}x)-\frac{u\rho_{e}-d\rho_{m}}{2c^2}(\alpha
x-\frac{1}{2\hbar\alpha}\theta
p_{y}))\hat{\textbf{e}}_{y}]^2-\frac{(u\rho_{e}-d\rho_{m})\hbar}{2mc^2},\end{array}\end{equation}
where we set $\Theta_{xy}=\theta$ and
$\bar{\Theta}_{xy}=\bar{\theta}$. After organization, the
Hamiltonian can be written as
\begin{equation}
\label{Eq:Rep.4}
\begin{array}{ll}
H=\frac{1}{2M}({p_x}^2+{p_y}^2)+\frac{1}{8}M\bar{\omega}^2(x^2+y^2)+\frac{\bar{\omega}}{2}(p_xy-p_yx)-\frac{(u\rho_{e}-d\rho_{m})\hbar}{2mc^2}
\end{array}\end{equation}
where the modified mass is
\begin{equation}
\label{Eq:Rep.4}
\begin{array}{ll}
M=m(\alpha+\frac{(u\rho_{e}-d\rho_{m})\theta}{4\alpha\hbar
c^2})^{-2}
\end{array}\end{equation}
and the modified cyclotron frequency is
\begin{equation}
\label{Eq:Rep.4}
\begin{array}{ll}
\bar{\omega}=\frac{2\sigma}{m}\mid(\alpha+\frac{(u\rho_{e}-d\rho_{m})\theta}{4\alpha\hbar
c^2})(\frac{\bar{\theta}}{2\alpha\hbar
}+\frac{(u\rho_{e}-d\rho_{m})\alpha}{2c^2})\mid=\frac{2}{m}(\alpha+\frac{(u\rho_{e}-d\rho_{m})\theta}{4\alpha\hbar
c^2})(\frac{\bar{\theta}}{2\alpha\hbar
}+\frac{(u\rho_{e}-d\rho_{m})\alpha}{2c^2}).
\end{array}\end{equation}
For the AC system and the HMW system, the modified masses and
cyclotron frequencies are
\begin{equation}
\label{Eq:Rep.4}
\begin{array}{ll}
M_{AC}=m(\alpha+\frac{u\rho_{e}\theta}{4\alpha\hbar c^2})^{-2},
\hspace{1cm}
\bar{\omega}_{AC}=\frac{2\sigma}{m}\mid(\alpha+\frac{u\rho_{e}\theta}{4\alpha\hbar
c^2})(\frac{\bar{\theta}}{2\alpha\hbar
}+\frac{u\rho_{e}\alpha}{2c^2})\mid=\frac{2}{m}(\alpha+\frac{u\rho_{e}\theta}{4\alpha\hbar
c^2})(\frac{\bar{\theta}}{2\alpha\hbar
}+\frac{u\rho_{e}\alpha}{2c^2})
\end{array}\end{equation}
and
\begin{equation}
\label{Eq:Rep.4}
\begin{array}{ll}
\hspace{1cm}M_{HMW}=m(\alpha-\frac{d\rho_{m}\theta}{4\alpha\hbar
c^2})^{-2}, \hspace{0.5cm}
\bar{\omega}_{HMW}=\frac{2\sigma}{m}\mid(\alpha-\frac{d\rho_{m}\theta}{4\alpha\hbar
c^2})(\frac{\bar{\theta}}{2\alpha\hbar
}-\frac{d\rho_{m}\alpha}{2c^2})\mid=\frac{2}{m}(\alpha-\frac{d\rho_{m}\theta}{4\alpha\hbar
c^2})(\frac{\bar{\theta}}{2\alpha\hbar
}-\frac{d\rho_{m}\alpha}{2c^2}).
\end{array}\end{equation}
Similarly we can define the creation and annihilation operators as
\begin{equation}
\label{Eq:Rep.4}
 \begin{array}{ll}
\bar{a}_x=\sqrt{{M|\bar{\omega}|}\over
8\hbar}(x+{{2\emph{i}p_x}\over
M|\bar{\omega}|})-i\sqrt{{M|\bar{\omega}|}\over
8\hbar}(y+{{2\emph{i}p_y}\over
M|\bar{\omega}|}),\hspace{2cm} \bar{a}^+_x=\sqrt{{M|\bar{\omega}|}\over 8\hbar}(x-{{2\emph{i}p_x}\over M|\bar{\omega}|})+i\sqrt{{M|\bar{\omega}|}\over 8\hbar}(y-{{2\emph{i}p_y}\over M|\bar{\omega}|}),                         \\~~\\
\bar{a}_y=i\sqrt{{M|\bar{\omega}|}\over
8\hbar}(x+{{2\emph{i}p_x}\over
M|\bar{\omega}|})-\sqrt{{M|\bar{\omega}|}\over
8\hbar}(y+{{2\emph{i}p_y}\over M|\bar{\omega}|}),\hspace{2cm}
\bar{\bar{a}}^+_y=-i\sqrt{{M|\bar{\omega}|}\over
8\hbar}(x-{{2\emph{i}p_x}\over
M|\bar{\omega}|})-\sqrt{{M|\bar{\omega}|}\over
8\hbar}(y-{{2\emph{i}p_y}\over M|\bar{\omega}|}).\\
\end{array}
\end{equation}
Thus, the Hamiltonian on NC phase space becomes the following form
\begin{equation}
\label{Eq:Rep.4}
 \begin{array}{ll}
H=\frac{\hbar|\bar{\omega}|}{2}({\bar{a}}^+_x{\bar{a}}_x+{\bar{a}}^+_y{\bar{a}}_y+1)+\frac{\hbar\sigma|\bar{\omega}|}{2}({\bar{a}}^+_y{\bar{a}}_y-{\bar{a}}^+_x{\bar{a}}_x)-\frac{(u\rho_{e}-d\rho_{m})\hbar}{2mc^2}.
\end{array}
\end{equation}
The energy eigenvalues and the corresponding eigenstates are
\begin{equation}
\label{Eq:Rep.4}
 \begin{array}{ll}
E_{\bar{n}_x,\bar{n}_y}=\frac{\hbar|\bar{\omega}|}{2}(\bar{n}_x+\bar{n}_y+1)+\frac{\hbar\sigma|\bar{\omega}|}{2}(\bar{n}_y-\bar{n}_x)-\frac{(u\rho_{e}-d\rho_{m})\hbar}{2mc^2}
\end{array}
\end{equation}
and
\begin{equation}
|{\bar{n}}_x,{\bar{n}}_y \rangle={1\over
\sqrt{{{\bar{n}}_x}!{{\bar{n}}_y!}}}({\bar{a}^+_x})^{{\bar{n}}_x}({\bar{a}^+_y})^{{\bar{n}}_y}|
0,0 \rangle.\label{h1}
\end{equation}
The levels for the AC system and the HMW system are
\begin{equation}
E^{AC}_{\bar{n}_x,\bar{n}_y}=\frac{\hbar|\bar{\omega}_{AC}|}{2}(\bar{n}_x+\bar{n}_y+1)+\frac{\hbar\sigma|\bar{\omega}_{AC}|}{2}(\bar{n}_y-\bar{n}_x)-\frac{u\rho_{e}\hbar}{2mc^2}
\end{equation}
and
\begin{equation}
E^{HMW}_{\bar{n}_x,\bar{n}_y}=\frac{\hbar|\bar{\omega}_{HMW}|}{2}(\bar{n}_x+\bar{n}_y+1)+\frac{\hbar\sigma|\bar{\omega}_{HMW}|}{2}(\bar{n}_y-\bar{n}_x)+\frac{d\rho_{m}\hbar}{2mc^2}.
\end{equation}
When $\bar{\theta}=0$, it leads $\alpha =1$\cite{Likang} and the
phase space becomes the NC space where only momentum-momentum is
commutative. The energy eigenvalues (24), (26) and (27) become
\begin{equation}
\label{Eq:Rep.4}
 \begin{array}{ll}
E_{\bar{n}'_x,\bar{n}'_y}=\frac{\hbar|\bar{\omega}'|}{2}(\bar{n}'_x+\bar{n}'_y+1)+\frac{\hbar\sigma|\bar{\omega}'|}{2}(\bar{n}'_y-\bar{n}'_x)-\frac{(u\rho_{e}-d\rho_{m})\hbar}{2mc^2},
\end{array}
\end{equation}
\begin{equation}
E^{AC}_{\bar{n}'_x,\bar{n}'_y}=\frac{\hbar|\bar{\omega}'_{AC}|}{2}(\bar{n}'_x+\bar{n}'_y+1)+\frac{\hbar\sigma|\bar{\omega}'_{AC}|}{2}(\bar{n}'_y-\bar{n}'_x)-\frac{u\rho_{e}\hbar}{2mc^2},
\end{equation}
\begin{equation}
E^{HMW}_{\bar{n}'_x,\bar{n}'_y}=\frac{\hbar|\bar{\omega}'_{HMW}|}{2}(\bar{n}'_x+\bar{n}'_y+1)+\frac{\hbar\sigma|\bar{\omega}'_{HMW}|}{2}(\bar{n}'_y-\bar{n}'_x)+\frac{d\rho_{m}\hbar}{2mc^2},
\end{equation}
respectively. Here
\begin{equation}
\bar{\omega}'=\frac{\sigma}{mc^2}|(1+\frac{(u\rho_{e}-d\rho_{m})\theta}{4\hbar
c^2})(u\rho_{e}-d\rho_{m})|=\frac{1}{mc^2}(1+\frac{(u\rho_{e}-d\rho_{m})\theta}{4\hbar
c^2})(u\rho_{e}-d\rho_{m}),
\end{equation}
\begin{equation}
\bar{\omega}'_{AC}=\frac{\sigma}{mc^2}|(1+\frac{u\rho_{e}\theta}{4\hbar
c^2})u\rho_{e}|=\frac{u\rho_{e}}{mc^2}(1+\frac{u\rho_{e}\theta}{4\hbar
c^2}),
\end{equation}
\begin{equation}
\bar{\omega}'_{HMW}=\frac{\sigma}{mc^2}|(1-\frac{d\rho_{m}\theta}{4\hbar
c^2})d\rho_{m}|=-\frac{d\rho_{m}}{mc^2}(1-\frac{d\rho_{m}\theta}{4\hbar
c^2}).
\end{equation}
When $\theta=\bar{\theta}=0$, the results return to the cases of
general quantum mechanics (7), (9) and (10).
\section{4. thermodynamics on NC situation}
Up to now, we can obtain Landau like levels of the Anandan system
for a given $\sigma$ on commutative space, NC space and NC phase
space. Glauber has discussed the coherent states of a harmonic
oscillator in detail.\cite{Glauber} Using the coherent state
method, we can know the quantum information of the system. Like
free electron gas related to Landau problems, we will study some
thermodynamical properties of such free atom gas with Anandan
interaction. The normalized coherent state of the system on NC
phase space is defined as
\begin{equation}
|z_x,z_y \rangle=Exp[-{1\over
2}(|z_x|^2+|z_y|^2)]Exp[z_x\bar{a}^+_x+z_y\bar{a}^+_y] |0,0
\rangle
\end{equation}
where $z_x$, $z_y$ are complex parameters. The mean coordinates of
the state $|z_x,z_y \rangle$ related to $\sigma$ is given by
\begin{equation}\label{Eq:Rep.4}
 \begin{array}{ll}
\bar{r}_{\sigma=-1}=(Re[Az_x^*(t)+B(-i z_y)], Im[Az_x^*(t)+B(-i
z_y)]) ,\\~~\\ \bar{r}_{\sigma=1}=(Re[Az_x^*+B(-i z_y(t))],
Im[Az_x^*+B(-i z_y(t))]),
\\
\end{array}\end{equation}
where $A=\alpha \sqrt{\frac{2\hbar}{M
{|\bar{\omega}|}}}-\frac{\theta}{2\alpha}\sqrt{\frac{M
{|\bar{\omega}|}}{2\hbar}}$, $B=\alpha \sqrt{\frac{2\hbar}{M
{|\bar{\omega}|}}}+\frac{\theta}{2\alpha}\sqrt{\frac{M{|\bar{\omega}|}}{2\hbar}}$
and $z_x^*(t)=z_x^*e^{i|\bar{\omega}|t}$,
$z_y(t)=z_ye^{-i|\bar{\omega}|t}$. Here we consider a
two-dimensional system with radius $R\gg A,B$. The partition
function of the system on NC phase space is
$Z=Tre^{-\frac{H}{kT}}$ in the standard way. For $\sigma=-1$ the
partition function can be written as
\begin{equation}
\label{Eq:Rep.4}
\begin{array}{ll}
Z=\frac{1}{\pi^2}\int d^2z_x d^2z_y \langle z_x,z_y
|e^{-\frac{1}{kT}[\frac{\hbar|\bar{\omega}|}{2}(2{\bar{a}}^+_x{\bar{a}}_x+1)-\frac{(u\rho_{e}-d\rho_{m})\hbar}{2mc^2}]}|z_x,z_y
\rangle\\~~\\
~~=\frac{4}{A^2}e^{-\frac{1}{kT}(\frac{\hbar|\bar{\omega}|}{2}-\frac{(u\rho_{e}-d\rho_{m})\hbar}{2mc^2})}
\int
|z'_x||z_y|Exp[|\frac{z'_x}{{A}}|^2(e^{-\frac{\hbar|\bar{\omega}|}{kT}}-1
)]d|z'_x|d|z_y|
\end{array}
\end{equation}
Like Ref.\cite{Albert}, we exclude the coherent states with the
mean coordinates outside the size and sum over the other states.
Because $R\gg A,B$ and the exponential falls off rapidly with
$|z'_y|$, we can integrate $|z'_y|$ from zero to infinity safely.
The result is
\begin{equation}
Z=\frac{R^2}{2Sinh[\frac{\hbar|\bar{\omega}|}{2kT}]}\frac{e^{\frac{(u\rho_{e}-d\rho_{m})\hbar}{2mc^2kT}}}{(\alpha
\sqrt{\frac{2\hbar}{M
{|\bar{\omega}|}}}+\frac{\theta}{2\alpha}\sqrt{\frac{M
{|\bar{\omega}|}}{2\hbar}})^2}. \label{h1}
\end{equation}
So for $\sigma=-1$ the free energy of the system is
\begin{equation}
\label{Eq:Rep.4}
\begin{array}{ll}
F=-nkTlnZ=-nkTln[\frac{R^2}{2Sinh[\frac{\hbar|\bar{\omega}|}{2kT}]}\frac{e^{\frac{(u\rho_{e}-d\rho_{m})\hbar}{2mc^2kT}}}{(\alpha
\sqrt{\frac{2\hbar}{M
{|\bar{\omega}|}}}+\frac{\theta}{2\alpha}\sqrt{\frac{M
{|\bar{\omega}|}}{2\hbar}})^2}].
\end{array}
\end{equation}
By using the same way, we can obtain the free energy for
$\sigma=1$
\begin{equation}
\label{Eq:Rep.4}
\begin{array}{ll}
F=-nkTln[\frac{R^2}{2Sinh[\frac{\hbar|\bar{\omega}|}{2kT}]}\frac{e^{\frac{(u\rho_{e}-d\rho_{m})\hbar}{2mc^2kT}}}{(\alpha
\sqrt{\frac{2\hbar}{M
{|\bar{\omega}|}}}-\frac{\theta}{2\alpha}\sqrt{\frac{M
{|\bar{\omega}|}}{2\hbar}})^2}].
\end{array}
\end{equation}
If we set $u\neq0$ and $d=0$, we can obtain the free energy of the
AC system on NC phase space,
\begin{equation}
\label{Eq:Rep.4}
\begin{array}{ll}
F=-nkTln[\frac{R^2}{2Sinh[\frac{\hbar|\bar{\omega}_{AC}|}{2kT}]}\frac{e^{\frac{u\rho_{e}\hbar}{2mc^2kT}}}{(\alpha
\sqrt{\frac{2\hbar}{M_{AC}
{|\bar{\omega}_{AC}|}}}+\frac{\theta}{2\alpha}\sqrt{\frac{M_{AC}
{|\bar{\omega}_{AC}|}}{2\hbar}})^2}] ~~~~~~ for~~\sigma=-1,
\end{array}
\end{equation}and
\begin{equation}
\label{Eq:Rep.4}
\begin{array}{ll}
F=-nkTln[\frac{R^2}{2Sinh[\frac{\hbar|\bar{\omega}_{AC}|}{2kT}]}\frac{e^{\frac{u\rho_{e}\hbar}{2mc^2kT}}}{(\alpha
\sqrt{\frac{2\hbar}{M_{AC}
{|\bar{\omega}_{AC}|}}}-\frac{\theta}{2\alpha}\sqrt{\frac{M_{AC}
{|\bar{\omega}_{AC}|}}{2\hbar}})^2}] ~~~~~~for~~ \sigma=1.
\end{array}
\end{equation}
If we set $u=0$ and $d\neq0$, we can obtain the free energy of the
HMW system on NC phase space,
\begin{equation}
\label{Eq:Rep.4}
\begin{array}{ll}
F=-nkTln[\frac{R^2}{2Sinh[\frac{\hbar|\bar{\omega}_{HMW}|}{2kT}]}\frac{e^{\frac{-d\rho_{m}\hbar}{2mc^2kT}}}{(\alpha
\sqrt{\frac{2\hbar}{M_{HMW}
{|\bar{\omega}_{HMW}|}}}+\frac{\theta}{2\alpha}\sqrt{\frac{M_{HMW}
{|\bar{\omega}_{HMW}|}}{2\hbar}})^2}] ~~~~~~ for~~\sigma=-1,
\end{array}
\end{equation}and
\begin{equation}
\label{Eq:Rep.4}
\begin{array}{ll}
F=-nkTln[\frac{R^2}{2Sinh[\frac{\hbar|\bar{\omega}_{HMW}|}{2kT}]}\frac{e^{\frac{-d\rho_{m}\hbar}{2mc^2kT}}}{(\alpha
\sqrt{\frac{2\hbar}{M_{HMW}
{|\bar{\omega}_{HMW}|}}}-\frac{\theta}{2\alpha}\sqrt{\frac{M_{HMW}
{|\bar{\omega}_{HMW}|}}{2\hbar}})^2}] ~~~~~~for~~ \sigma=1.
\end{array}
\end{equation}
Further, by restricting $\theta$, $\bar{\theta}$, we can obtain
the free energies on NC space and commutative space for the
Anandan system, the AC system and the HMW system.

\textbf{(1). High Temperature Approximation}\\

For the Anandan system where $|u\rho_{e}-d\rho_{m}|$ is not very
small, the free energy for $\sigma=-1$ on NC phase space can be
expressed by
\begin{equation}
F_{-1}=F_{-1}^0+F_{-1}^\theta+F_{-1}^{\bar{\theta}}\\
\end{equation}
where
\begin{equation}
\label{Eq:Rep.4}
\begin{array}{ll}
F_{-1}^0=-nkTln[\frac{mR^2|\omega|e^{-\frac{\hbar|\omega|}{2kT}}Csch[\frac{\hbar|\omega|\alpha^2}{2kT}]}{4\hbar\alpha^2}],\\~~~\\
F_{-1}^\theta=\frac{kmnT|\omega|\theta}{4\hbar\alpha^2}-\frac{1}{8}Coth[\frac{\hbar|\omega|\alpha^2}{2kT}]mn|\omega|^2\theta,\\~~\\
F_{-1}^{\bar{\theta}}=\frac{knT\bar{\theta}}{\hbar
m|\omega|\alpha^2}-\frac{Coth[\frac{\hbar|\omega|\alpha^2}{2kT}]n
\bar{\theta}}{2m}.\\
\end{array}
\end{equation}
In the calculation, we only consider the first order modification
from space-space non-commutativity and momentum-momentum
non-commutativity and ignore the interaction between them.
Similarly, we can obtain for $\sigma=1$
\begin{equation}
F_{1}=F_{1}^0+F_{1}^\theta+F_{1}^{\bar{\theta}}\\
\end{equation}
where
\begin{equation}
\label{Eq:Rep.4}
\begin{array}{ll}
F_{1}^0=-nkTln[\frac{mR^2|\omega|e^{\frac{\hbar|\omega|}{2kT}}Csch[\frac{\hbar|\omega|\alpha^2}{2kT}]}{4\hbar\alpha^2}],\\~~~\\
F_{1}^\theta=-\frac{kmnT|\omega|\theta}{4\hbar\alpha^2}+\frac{1}{8}Coth[\frac{\hbar|\omega|\alpha^2}{2kT}]mn|\omega|^2\theta,\\~~\\
F_{1}^{\bar{\theta}}=-\frac{knT\bar{\theta}}{\hbar
m|\omega|\alpha^2}+\frac{Coth[\frac{\hbar|\omega|\alpha^2}{2kT}]n
\bar{\theta}}{2m}.\\
\end{array}
\end{equation}
In high temperature approximation, the free energy on NC phase
space is
\begin{equation}
\label{Eq:Rep.4}
\begin{array}{ll}
F_{-1}=Tkn(ln[\frac{1}{T}]-ln[\frac{kmR^2}{2\hbar^2\alpha^4}])+\frac{\hbar
n|\omega|}{2}+\frac{1}{T}(\frac{n\hbar^2|\omega|^2\alpha^4}{24k}-\frac{\hbar
mn|\omega|^3\alpha^2\theta}{48k}-\frac{n\hbar|\omega|\alpha^2\bar{\theta}}{12km}),\\~~\\
F_{1}=Tkn(ln[\frac{1}{T}]-ln[\frac{kmR^2}{2\hbar^2\alpha^4}])-\frac{\hbar
n|\omega|}{2}+\frac{1}{T}(\frac{n\hbar^2|\omega|^2\alpha^4}{24k}+\frac{\hbar
mn|\omega|^3\alpha^2\theta}{48k}+\frac{n\hbar|\omega|\alpha^2\bar{\theta}}{12km}).
\end{array}
\end{equation}
Further, if we restrict the values $\theta$, $\bar{\theta}$ and
$\alpha$, we can get the free energies on NC space and commutative
space, respectively,
\begin{equation}
\label{Eq:Rep.4}
\begin{array}{ll}
F_{-1}^{NC
~Space}=Tkn(ln[\frac{1}{T}]-ln[\frac{kmR^2}{2\hbar^2}])+\frac{\hbar
n|\omega|}{2}+\frac{1}{T}(\frac{n\hbar^2|\omega|^2}{24k}-\frac{\hbar
mn|\omega|^3\theta}{48k}),\\~~\\
F_{1}^{NC
~Space}=Tkn(ln[\frac{1}{T}]-ln[\frac{kmR^2}{2\hbar^2}])-\frac{\hbar
n|\omega|}{2}+\frac{1}{T}(\frac{n\hbar^2|\omega|^2}{24k}+\frac{\hbar
mn|\omega|^3\theta}{48k})
\end{array}
\end{equation}
and
\begin{equation}
\label{Eq:Rep.4}
\begin{array}{ll}
F_{-1}^{C
~Space}=Tkn(ln[\frac{1}{T}]-ln[\frac{kmR^2}{2\hbar^2}])+\frac{\hbar
n|\omega|}{2}+\frac{1}{T}\frac{n\hbar^2|\omega|^2}{24k},\\~~\\
F_{1}^{C
~Space}=Tkn(ln[\frac{1}{T}]-ln[\frac{kmR^2}{2\hbar^2}])-\frac{\hbar
n|\omega|}{2}+\frac{1}{T}\frac{n\hbar^2|\omega|^2}{24k}
\end{array}
\end{equation}

\textbf{(2). Zero Temperature Limit}\\

According to Eqs. (38)and(39), if we consider $T\rightarrow0$, we
will find the free energy on NC phase space is
\begin{equation}
\label{Eq:Rep.4}
\begin{array}{ll}
F_{-1}\rightarrow\frac{\hbar n|\omega|}{2}+\frac{\hbar
n|\omega|\alpha^2}{2}-\frac{1}{8}mn|\omega|^2\theta-\frac{n\bar{\theta}}{2m},\\~~\\
F_{1}\rightarrow-\frac{\hbar n|\omega|}{2}+\frac{\hbar
n|\omega|\alpha^2}{2}+\frac{1}{8}mn|\omega|^2\theta+\frac{n\bar{\theta}}{2m}.
\end{array}
\end{equation}
The results on NC space and commutative space are
\begin{equation}
\label{Eq:Rep.4}
\begin{array}{ll}
F_{-1}^{NC
~Space}\rightarrow\hbar n|\omega|-\frac{1}{8}mn|\omega|^2\theta,\\~~\\
F_{1}^{NC ~Space}\rightarrow\frac{1}{8}mn|\omega|^2\theta
\end{array}
\end{equation}
and
\begin{equation}
\label{Eq:Rep.4}
\begin{array}{ll}
F_{-1}^{C
~Space}\rightarrow\hbar n|\omega|,\\~~\\
F_{1}^{C ~Space}\rightarrow0.
\end{array}
\end{equation}
Here we emphasize the two points for the two particular cases of
the temperature. (i)The free energy of the Anandan system can be
restricted to the expressions of the AC system and the HMW system
by making the replacement $|\omega|\rightarrow|\omega_{AC}|$ and
$|\omega|\rightarrow|\omega_{HMW}|$ respectively; (ii) For a given
$\sigma$, the free energies have some difference among NC phase
space, NC space and commutative space which may provide some clues
to verify the presence of NC situation in future. For example, the
free energy tends to zero for $\sigma=1$ on commutative space in
zero temperature limit, but this phenomenon doesn't occur on NC
situation. \\
\section{5. revolution direction and Maxwell duality}
As we know, a coherent state represents a way that is as close  as
possible to classical localization. In the sections above. we
introduced the sign $\sigma$ which describes the revolution
direction  of the corresponding classical motion in fact. Here we
use the most classical quantum state, the coherent state, to
clearly catch the revolution direction $\sigma$. The mean
coordinates of the state $|z_x,z_y \rangle$ related to $\sigma$ is
given by Eq.(35). For $\sigma=-1$, the wave packet centroid of the
coherent state $|z_x,z_y \rangle$ moves anticlockwise with the
radius $|Az_x^*|$ and the frequency $|\bar{\omega}|$. For
$\sigma=1$, it moves clockwise with the radius  $|Bz_y|$and the
same frequency $|\bar{\omega}|$. So here $\sigma=-1$ describes
anticlockwise revolution and $\sigma=1$ describes clockwise
revolution, which is all right for the Anandan system, the AC
system and the HMW system on commutative space, NC
space and NC phase space.\\

On any given space among commutative space, NC space and NC phase
space, for a given revolution direction, Landau like levels of the
Anandan system are invariant and the levels between the AC system
and the HMW system become each other under Maxwell dual
transformations,
\begin{equation}
\label{Eq:Rep.4}
\begin{array}{ll}
\rho_e\rightarrow\rho_m,~~~d\rightarrow u,\\
\rho_m\rightarrow-\rho_e,~~~u\rightarrow -d.
\end{array}
\end{equation}
This result can be explained by the invariance of the Anandan
Hamiltonian and the interchangeablity between the AC Hamiltonian
and the HMW Hamiltonian under Maxwell dual transformations on the
corresponding space. Thus Landau like levels of the AC system and
the HMW system are similar under the same revolution direction. In
Ref.\cite{Ribeiro}, Ribeiro \emph{et al.} think that Landau like
levels of the HMW system have the same form as the levels of the
AC system with the opposite direction. But that's not true. For
example, in terms of their paper, the AC system ($u\rho_e>0$)  and
the HMW system ($d\rho_m<0$) should have the same form of the
levels and opposite directions. But in fact the cases of
$u\rho_e>0$ and $d\rho_m<0$ describe the same revolution direction
as shown in the movement of the wave packet centroid of the
coherent state $|z_x,z_y \rangle$ for $\sigma=1$ above. The reason
of the mistake is that the definition of the cyclotron frequencies
in their paper break Maxwell duality. However, in our paper, the
cyclotron frequencies $\omega_{AC}=\sigma\frac{\mid
u\rho_{e}\mid}{mc^2}=\frac{ u\rho_{e}}{mc^2}$ and
$\omega_{HMW}=\sigma\frac{\mid d\rho_{m}\mid}{mc^2}=\frac{
-d\rho_{m}}{mc^2}$ will change each
other under Maxwell dual transformations.\\
\section{6. supersymmetry}
Now, we have known landau like levels of the Anandan system on
commutative space, NC space and NC phase space can be all divided
into two classes labelled by  the revolution direction $\sigma$.
In this section we will utilize supersymmetry to study the
difference of the system between commutative space and NC
situation by transforming the Hamiltonian (16) again. We introduce
the new annihilation and creation operators
\begin{equation}
\label{Eq:Rep.4}
\begin{array}{ll}
\hat{b'}=\frac{1}{\sqrt{2M\hbar|\bar{\omega}|}}(\prod_x+i\sigma\prod_y),\\
\hat{b'}^+=\frac{1}{\sqrt{2M\hbar|\bar{\omega}|}}(\prod_x-i\sigma\prod_y),
\end{array}
\end{equation}
where
$\prod_x=p_x+\frac{2c^2\bar{\theta}+2\alpha^2\hbar(u\rho_{e}-d\rho_{m})}{4\alpha^2\hbar
c^2+(u\rho_{e}-d\rho_{m})\theta}y$ and
$\prod_y=p_y-\frac{2c^2\bar{\theta}+2\alpha^2\hbar(u\rho_{e}-d\rho_{m})}{4\alpha^2\hbar
c^2+(u\rho_{e}-d\rho_{m})\theta}x$. Thus the Hamiltonian (16) can
be expressed as
\begin{equation}
\label{Eq:Rep.4}
\begin{array}{ll}
H=\hbar|\bar{\omega}|(\hat{b'}^+\hat{b'}+\frac{1}{2})-\frac{\hbar\sigma|\omega|}{2}.
\end{array}
\end{equation}
On commutative space, $|\bar{\omega}|$ becomes $|\omega|$, and
$\hat{b'}$ becomes $\hat{b}$ where
$\hat{b}=\frac{1}{\sqrt{2m\hbar|\omega|}}(p_x+\frac{u\rho_{e}-d\rho_{m}}{2
c^2}y+i\sigma p_y-i\sigma\frac{u\rho_{e}-d\rho_{m}}{2 c^2}x)$. So
the Hamiltonian on this space is
\begin{equation}
\label{Eq:Rep.4}
\begin{array}{ll}
H_C=\hbar|\omega|(\hat{b}^+\hat{b}+\frac{1}{2}-\frac{\sigma}{2}).
\end{array}
\end{equation}
Here we introduce fermionic annihilation and creation operators
$\hat{d}$, $\hat{d}^+$ and suppose the eigenvalue of
$[\hat{d},\hat{d}^+]$ is $\sigma$. Thus the Hamiltonian $H_C$ can
be expressed as
$H_C=\hbar|\omega|(\hat{b}^+\hat{b}+\hat{d}^+\hat{d})$. The
supercharge can be defined as
\begin{equation}
Q=\sqrt{\hbar|\omega|}\hat{b}\hat{d}^+
\end{equation}
which together with the Hamiltonian $H_C$ close the surperalgebra
\begin{equation}
\label{Eq:Rep.4}
\begin{array}{ll}
Q^2=(Q^+)^2=0,~~~~~~~H_C=\{Q,~Q^+\}.
\end{array}
\end{equation}
The extended Fock states can be defined as $|n_b, n_d,
\kappa\rangle$ where $n_b$, $n_d$ are the eigenvalues of the
number operators $\hat{b}^+\hat{b}$, $\hat{d}^+\hat{d}$ and
$\kappa$ is a good quantum number. These eigenstates can be
related by the good supersymmetry transformation
\begin{equation}
\label{Eq:Rep.4}
\begin{array}{ll}
|n_b-1, 1, \kappa\rangle=\frac{1}{\sqrt{E_{n_b}}}Q|n_b, 0,
\kappa\rangle,~~~~~~~~~~~~~~~|n_b+1, 0,
\kappa\rangle=\frac{1}{\sqrt{E_{n_b+1}}}Q^+|n_b, 1,
\kappa\rangle\\
\end{array}
\end{equation}
and
\begin{equation}
\label{Eq:Rep.4}
\begin{array}{ll}
Q|0, 0, \kappa\rangle=Q^+|0, 0, \kappa\rangle=0,
\end{array}
\end{equation}
where $E_{n_b}$ is the energy eigenvalues of the states $|n_b, 0,
\kappa\rangle$, $|n_b-1, 1, \kappa\rangle$ and $E_{n_b+1}$ is the
ones of the states $|n_b, 1, \kappa\rangle$, $|n_b+1, 0,
\kappa\rangle$. According to the Hamiltonian (56), the energy
eigenvalues on NC phase space is
\begin{equation}
E_{n',\sigma}=\hbar|\bar{\omega}|(n'+\frac{1}{2})-\frac{\hbar\sigma|\omega|}{2}
\end{equation}
where $n'$ is the eigenvalue of the number operator
$\hat{b'}^+\hat{b'}$. And the corresponding eigenstates are
$|n',\kappa' \rangle_\sigma$ where $\kappa'$ is a good quantum
number. Obviously, for different $\sigma$, the energy eigenvalues
are not equal, that is, $E_{n'_1,1}\neq E_{n'_2,-1}$. Thus the two
sets of eigenstates labelled by $\sigma$ can not be related by a
supersymmetry transformation on NC phase space. The phenomenon
also exists on NC space. The difference resulted from
supersymmetry between commutative space and NC situation also
occurs for the AC system and the HMW system.\\
\section{6. Conclusion}
In this letter we study the Anandan system in a special
electromagnetic field, where we find unlike the cases of the AC
system and the HMW system the torques on the magnetic dipole and
the electric dipole don't vanish. We obtain the Landau analog
levels and eigenstates on commutative space, NC space and NC phase
space. We study some statistical properties of such free atom gas
with the Anandan interaction and present the expressions of some
thermodynamic quantities related to revolution direction. Some
simple formulae of the free energy in two cases of temperature on
the three spaces are also obtained. Some difference of the free
energy among NC phase space, NC space and commutative space may
provide us some clues to verify the presence of NC situation in
future. The relation between the value of $\sigma$ and revolution
direction is presented clearly. We find Landau like levels of the
Anandan system are invariant and the levels between the AC system
and the HMW system become each other under Maxwell dual
transformations on the three spaces. And point the mistake of
Ribeiro \emph{et al.}, Landau like levels of the HMW system have
the same form as the levels of the AC system with the opposite
direction. We find the two sets of eigenstates labelled by
$\sigma$ can be related by a supersymmetry transformation on
commutative space, but the phenomenon don't occur on NC situation.
Some results relevant to the AC system and the HMW system are also
obtained by restricting the magnetic dipole and the electric
dipole.\\

\textbf{Acknowledgments:} This work was supported by the National Natural Science Foundation of China (10875035),
 and  Natural Science Foundation of Zhejiang Province (Y6110470),


\begin{thebibliography}{99}

\bibitem{AB}
Y. Aharonov, D. Bohm, Phys. Rev. 115, 485 (1959).
\bibitem{AC}
Y Aharonov and A. Casher, Phys. Rev. Lett. 53, 319, (1984)

\bibitem{Badurek}
G. Badurek, H. Weinfurter, R. G\"{a}hler, A. Kollmar, S. Wehinger,
and A. Zeilinger, Phys. Rev. Lett. 71, 307 (1993); M. Peshkin and
H. J. Lipkin, Phys. Rev. Lett. 74, 2847 (1995)
\bibitem{Cimmino}
A. Cimmino et al., Phys. Rev. Lett. 63, 380 (1989).
\bibitem{Sangster}
K. Sangster et al., Phys. Rev. Lett. 71, 3641 (1993); K. Sangster
et al., Phys. Rev. A 51, 1776 (1995).
\bibitem{HMW}
X. G. He and B. H. J. McKellar, Phys. Rev. A, 47, 3424 (1993); M.
Wilkens, Phys. Rev. Lett. 72, 5 (1994).
\bibitem{WHW}
H. Wei, R. Han and X. Wei, Phys. Rev. Lett. 75, 2071 (1995).
\bibitem{Dowling}
J.P. Dowling, C.P.Willian, J.D. Franson, Phys. Rev. Lett. 83, 2486
(1999)
\bibitem{AnandanR}
J. Anandan, Phys. Rev. Lett. 85, 1354 (2000).
\bibitem{AnandanL}
J. Anandan, Phys. Lett. A 138, 347 (1989).
\bibitem{Furtado}
C. Furtado and C. A. de Lima Ribeiro, Phys. Rev A 69, 064104
(2004).
\bibitem{Paredes}
B. Paredes, P. Fedichev, J. I. Cirac, and P. Zoller, Phys. Rev.
Lett. 87, 010402 (2001).

\bibitem{Ericsson}
M. Ericsson, E. Sj$\ddot{o}$qvist, Phys. Rev. A 65, 013607 (2001)
\bibitem{Ribeiro}
L.R. Ribeiro, C. Furtado, J.R. Nascimento, Phys. Lett. A 348, 135
(2006)

\bibitem{SW}
N. Seiberg and E. Witten, JHEP 9909: 032(1999), hep-th/9908142.

\bibitem{CDS}
A. Connes, M. R. Douglas, A. Schwarz, JHEP  9802, 003(1998),
hep-th/9711162; M. R. Douglas, C. M. Hull, JHEP 9802, 008(1998),
hep-th/9711165; M. R. Douglas and N. A. Nekrasov, Rev. Mod. Phys.
73, 977(2001).

\bibitem{AAS}
F. Ardalan, H. Arfaei, M. M. Sheikh-Jabbari, JHEP  9902,
016(1999), hep-th/9810072.

\bibitem{CFZ} T. Curtright, D. Fairlie and C. Zachos,
Phys. Rev. D58 025002 (1998); L. Mezincescu, hep-th/0007046.

\bibitem{CH}
S-C. Chu, P-M. Ho, Nucl. Phys. B550, 151(1999), hep-th/9812219;
 Nucl. Phys. B568, 447(2000), hep-th/9906192.

\bibitem{Scho}
V. Schomerus, JHEP  9906, 030(1999),  hep-th/9903205.

\bibitem{Thom}
Thomas Filk, Phys. Lett. B376, 53(1996).

\bibitem{Leon}
Leonardo Castellani, Class. Quant. Grav. 17, 3377(2000),
hep-th/0005210.

\bibitem{Kone}
A. Konechny and A. Schwarz, Phys. Rept. 360, 353(2002).
\bibitem{Dou}
M. R. Douglas and N. A. Nekrasov, Rev. Mod. Phys. 73, 977(2001).

\bibitem{nair} V. P. Nair, A. P. Polychronakos,
Phys. Lett. B505, 267(2001); J. Z. Zhang, Phys. Lett. B 584, 204
(2004)
\bibitem{Chaichian1}
M. Chaichian, M.M. Sheikh-Jabbari, A. Tureanu, Phys. Rev. Lett.
86, 2716 (2001).
\bibitem{Chaichian2}
M. Chaichian, A. Demichev, P. Presnajder, M.M. Sheikh-Jabbari, A.
Tureanu, Nucl. Phys. B 611, 383 (2001).
\bibitem{Chaichia3}
M. Chaichian, P. Presnajder, M.M. Sheikh-Jabbari, A. Tureanu,
Phys. Lett. B 527, 149 (2002).
\bibitem{Falomir}
 H. Falomir, J. Gamboa, M. Loeve, F.
Mendez, J.C. Rojas, Phys. Rev. D 66, 045018 (2002).

\bibitem{klisd}
K. Li and S. Dulat, Eur. Phys. J. C 46, 825 (2006).







\bibitem{Mirza}
B. Mirza and M. Zarei, Eur. Phys. J. C 32, 583 (2004).
\bibitem{Mirza1}
 S. Dulat and K. Li, Eur. Phys. J. C 54, 333 (2008).

\bibitem{Dulat}
S. Dulat, K. Li and Jianhua Wang, J. PHYS. A: MATH. THEOR. 41,
065303 (2008).


\bibitem{Horvathy1}
P.A. Horvathy, Ann. Phys. (N.Y.) 299, 128 (2002).
\bibitem{Horvathy2}
P.A. Horvathy, M.S. Plyushchay, Nucl. Phys. B 714, 269 (2005).
\bibitem{Gamboa}
 J. Gamboa, M. Loewe, F. Mendez, J.C. Rojas, Mod. Phys. Lett. A 16,
2075 (2001).

\bibitem{Susskind}
L. Susskind, arXiv:hep-th/0101029.
\bibitem{Dayi}
O.F. Dayi, A. Jellal, J. Math. Phys. 43, 4592 (2002).
\bibitem{Basu}
B. Basu, S. Ghosh, Phys. Lett. A 346, 133 (2005)



\bibitem{Passos}
E. Passos, L. R. Ribeiro, C. Furtado and J. R. Nascimento Phys.
Rev. A 76, 012113 (2007)
\bibitem{RibeiroE}
L.R. Ribeiro, E. Passos C, Furtado and J.R. Nascimento Eur. Phys.
J. C  56, 597 (2008)
\bibitem{Likang} Kang Li, Jianhua Wang, Chiyi Chen, Mod. Phys. Lett.
A20, 2165(2005).
\bibitem{Glauber}
R. J. Glauber, Phys. Rev. 131, 2766 (1963)
\bibitem{Albert} Albert Feldman and Arnold H.Kahn, Phys. Rev. B1 4584(1970)





\end{thebibliography}
\end{document}